\documentclass[10pt,american,prl,floatfix,superscriptaddress,twocolumn,showpacs,amsmath,amssymb,reprint,nobibnotes,nofootinbib]{revtex4-1}
\usepackage[T1]{fontenc}
\usepackage[latin9]{inputenc}
\usepackage{pslatex}
\usepackage{color}
\usepackage{babel}
\usepackage{amsmath}
\usepackage{graphicx}
\usepackage{esint}
\usepackage[unicode=true,pdfusetitle,
 bookmarks=true,bookmarksnumbered=false,bookmarksopen=false,
 breaklinks=true,pdfborder={0 0 0},backref=false,colorlinks=true]
 {hyperref}

\makeatletter

\providecommand{\tabularnewline}{\\}

\PassOptionsToPackage{caption=false}{subfig} 
\usepackage{hyperref}
\hypersetup{	
breaklinks=true,
colorlinks=true,
citecolor=blue,
linkcolor=blue,
filecolor=blue,
urlcolor=blue
}
\makeatother

\usepackage{babel}
\begin{document}
\title{Emergent SU($N$) symmetry in disordered SO($N$) spin chains}
\author{V. L. Quito}
\affiliation{Department of Physics and Astronomy, Iowa State University, Ames,
Iowa 50011, USA}
\affiliation{Department of Physics and National High Magnetic Field Laboratory,
Florida State University, Tallahassee, Florida 32306, USA }
\author{P. L. S. Lopes}
\affiliation{Stewart Blusson Quantum Matter Institute, University of British Columbia,
Vancouver, British Columbia, Canada V6T 1Z4}
\affiliation{D\'{e}partement de Physique, Institut Quantique and Regroupement
Qu\'{e}b\'{e}cois sur les Mat\'{e}riaux de Pointe, Universit\'{e}
de Sherbrooke, Sherbrooke, Qu\'{e}bec, Canada J1K 2R1}
\author{Jos\'{e} A. Hoyos}
\affiliation{Instituto de F\'{i}sica de S\~{a}o Carlos, Universidade de S\~{a}o
Paulo, C.P. 369, S\~{a}o Carlos, SP 13560-970, Brazil}
\author{E. Miranda}
\affiliation{Instituto de F\'{i}sica Gleb Wataghin, Unicamp, Rua S\'{e}rgio Buarque
de Holanda, 777, CEP 13083-859 Campinas, SP, Brazil}
\date{\today}
\begin{abstract}
Strongly disordered spin chains invariant under the SO($N$) group are
shown to display random-singlet phases with emergent SU($N$) symmetry
without fine tuning. The phases with emergent SU($N$) symmetry are of
two kinds: one has a ground state formed of randomly distributed singlets
of strongly bound pairs of SO($N$) spins (a `mesonic' phase), while
the other has a ground state composed of singlets made out of strongly
bound integer multiples of $N$ SO($N$) spins (a `baryonic' phase). The
established mechanism is general and we put forward the cases of $N=2,3,4$
and $6$ as prime candidates for experimental realizations in material
compounds and cold-atoms systems. We display universal
temperature scaling and critical exponents for susceptibilities distinguishing
these phases and characterizing the enlarging of the microscopic symmetries
at low energies.
\end{abstract}
\maketitle
\emph{Introduction.}\textendash{} The process of symmetry breaking,
as the energy of a given system is lowered, plays a central role in
our current understanding of both high-energy physics (as in the standard
model) and condensed matter physics (with universality and classification
of phases) \cite{AndersonBook,Laughlin04012000}. A less noticed
(and explored) scenario is that of symmetry emergence, in which the
lowering of the system's energy allows for ground states and excitations
which are symmetric under a larger group of transformations than their
corresponding microscopic Hamiltonian. A basic mechanism by which
this can happen can be understood in the renormalization group framework
by means of fixed points characterized by a symmetry which is broken
only by irrelevant perturbations. There remains, nevertheless, a widespread
lack of recognizable generic processes or patterns, so systems which
realize this type of physics are found by trial and error (see Refs.~\onlinecite{zamolodchikov89,PhysRevLett.115.166401,schmalianbatista08,Coldeaetal2010,Damle2002,Senthietal2004,Groveretal2014,batistaortiz04,fidkowski-etal-prb09,Yipetal2015,zohar2016,PhysRevB.55.8295,linetal98}
for examples). In scenarios dominated by disorder, the situation is
even more clouded. It was in this context that, in Ref.~\onlinecite{Quito_PhysRevLett.115.167201},
it was shown that generic disordered SU(2)-symmetric spin-1 chains
exhibit emergent SU(3)-symmetric random-singlet phases (RSPs)~\cite{madasguptahu}.
There, it was also noted that in the pioneering work by Fisher on
disordered XXZ spin-1/2 chains~\cite{fisher94-xxz}, there was also
the emergence of SU(2) symmetric RSPs; SU(2) is explicitly broken
down to U(1) in the microscopic XXZ Hamiltonian. What was \emph{not}
noted, however, is that in \textit{both} cases the emergent SU($N$)
symmetry materialized out of systems with manifest SO($N$) invariance,
with $N=3$ and $2$, respectively. 

This situation, which at first might be naively thought of as just
a coincidence, uncovers, on the contrary, a consistent pattern. It
is the aim of this Letter to show that generic disordered magnetic
chains invariant under the SO($N$) group, in its defining vector representation,
display emergent SU($N$)-symmetric phases via a unified route for any
$N\geq2$; we denote this process by $\text{SO}(N)\overset{\textrm{emerg}}{\longrightarrow}\text{SU}(N)$.
Our pattern of symmetry emergence contains two phases: (i) an obvious
SU($N$) generalization of the SU(2)-symmetric random singlet phase of
the Heisenberg chain of Ref.~\onlinecite{fisher94-xxz}, (ii) a phase
whose ground state also consists of random SU($N$)-symmetric singlets,
but which are composed of $kN$ original SO($N$) `spins', with $k$
an arbitrary integer. Separating the two phases there is a critical
point with manifest SU($N$) symmetry. In the particular case of $\text{SO}(3)\overset{\textrm{emerg}}{\longrightarrow}\text{SU}(3)$
of Ref.~\onlinecite{Quito_PhysRevLett.115.167201} {[}previously interpreted
as $\text{SU}(2)_{\textrm{spin-1}}\overset{\textrm{emerg}}{\longrightarrow}\text{SU}(3)${]},
particular versions of these phases were dubbed ``mesonic'' and
``baryonic'' random singlet phases, respectively. Furthermore, every
one-dimensional RSP encountered so far~\cite{PhysRevB.70.180401,Bonesteel2007,fidkowski-etal-prb08,fidkowski-etal-prb09}
seems to find a counterpart in one of the permutation-symmetric multicritical
points described by Damle and Huse~\cite{PhysRevB.66.104425,Damle2002},
each one indexed by an integer $n$. The SO($N$) baryonic RSPs we found
realize all of these Damle-Huse points with $n=N$ in an
extended phase (see also the discussion in Ref.~\onlinecite{QuitoLopes2016PRB})

While SO($N$) magnetism may sound exotic at first, such systems can
be realized in several ways, either by exploiting explicit breaking
of a larger SU($N$) isotropy or, more interestingly, by taking advantage
of the isomorphisms between orthogonal (so($N$))
and unitary (su($N$)) algebras at low $N$ values. 
Some examples, summarized in Table~\ref{tab:Summary-known-models},
follow:

(i) The first two mentioned cases, that of the XXZ spin-1/2 Heisenberg
chain~\cite{fisher94-xxz} and of spin-1 bilinear and biquadratic
Hamiltonians~\cite{Quito_PhysRevLett.115.167201} can be realized
in solid state~\cite{Toskovic2016} and, in principle, in cold atom
systems~\cite{PhysRevLett.93.250405,PhysRevA.68.063602}, respectively.
The former has a Hamiltonian with broken SU(2)-symmetry which, in
fact, corresponds to an SO(2) symmetric Hamiltonian. The latter is
realized explicitly as the most general SU(2)-symmetric Hamiltonian
with spin-1 representations, but due to the algebra isomorphism $\mathrm{so}(3)\sim\mathrm{su}(2)$,
it corresponds also to the most general SO(3)-symmetric Hamiltonian
in the defining vector representation.

(ii) Through the isomorphism $\mathrm{so}(4)\sim\mathrm{su}(2)\otimes\mathrm{su}(2)$,
SO(4)-symmetric magnetism is realized by the well-known Kugel\textendash Khomskii
Hamiltonian~\cite{0038-5670-25-4-R03}, commonly used in the description
of $e_{g}$ orbitals in transition metal oxides \cite{Tokura_2000},
with $\mathrm{su}\left(2\right)$-spin ($\mathbf{S}$) and $\mathrm{su}\left(2\right)$-orbital
($\mathbf{T}$) degrees of freedom
\begin{eqnarray}
H_{KK} & = & \sum_{i}\left[J_{i}\left(\mathbf{S}_{i}\cdot\mathbf{S}_{i+1}+\mathbf{T}_{i}\cdot\mathbf{T}_{i+1}\right)\right.\nonumber \\
 &  & +\left.8D_{i}\left(\mathbf{S}_{i}\cdot\mathbf{S}_{i+1}\right)\left(\mathbf{T}_{i}\cdot\mathbf{T}_{i+1}\right)\right].\label{eq:KK}
\end{eqnarray}

(iii) There are proposals to realize SU($N$) magnetism with arbitrary
$N$ in fermionic alkaline-earth cold atomic systems in representations
other than the fundamental one~\cite{Gorshkov2010}. Exploiting
the isomorphism $\mathrm{so}\left(6\right)\sim\mathrm{su}\left(4\right)$,
disordered SU(4) magnetic chains in the self-conjugate representation
realize an SO(6)-symmetric chain in its defining representation. In
this case, according to our mechanism, disordered SU(4) symmetric
chains would realize $\ensuremath{\text{SU\ensuremath{\left(4\right)}}\overset{\mbox{emerg}}{\longrightarrow}\text{SU\ensuremath{\left(6\right)}}}$.~\footnote{The proposal from Ref.~\onlinecite{Gorshkov2010} generates SU($N$)-symmetric
spin Hamiltonians, with arbitrary $N$, in perturbation theory in $1/U$
(where $U$ is the usual Hubbard on-site interaction) in the Mott
insulating limit. To lowest order, only the Heisenberg term appears.
By symmetry, however, other SU($N$)-invariant terms (biquadratic, etc.)
are also allowed and appear in higher orders of perturbation theory.
Systems which are closer to the Mott transition and at weaker interaction
strengths should therefore be described by these more general SU($N$)
Hamiltonians.}

(iv) Random SO($2S+1$) chains can, in fact, be designed by fine-tuning
in \emph{any} disordered rotation invariant spin-$S$ system. Such generic
spin-$S$ chains have been previously studied by some of us~\cite{quitoprb2016},
but the SO($N$) phases of these systems were not characterized at that point.

\begin{table}
\begin{centering}
\begin{tabular}{ccc}
$\begin{array}{c}
\text{\ensuremath{\mathrm{Hamiltonian}}}\\
\mathrm{symmetry}
\end{array}$ & Possible realizations & $\begin{array}{c}
\text{Emergent}\\
\text{symmetry}
\end{array}$\tabularnewline
\hline 
\hline 
SO(2) & anisotropic spin-1/2 systems & SU(2)\tabularnewline
SO(3) & generic spin-1 systems & SU(3)\tabularnewline
SO(4) & $e_{g}$ orbitals in transition metal oxides & SU(4)\tabularnewline
SO(6) & cold fermionic alkaline-earth atoms & SU(6)\tabularnewline
\end{tabular}
\par\end{centering}
\caption{List of the most relevant SO($N$)-symmetric one-dimensional models described
by Eq.~\eqref{eq:Hamilt} with their possible physical realizations
and the corresponding emergent symmetry in the limit of strong disorder.
\label{tab:Summary-known-models}}
\end{table}

We will first describe the general model and our results for the disordered
$\text{SO}(N)\overset{\textrm{emerg}}{\longrightarrow}\text{SU}(N)$
mechanism. After that, we will give the finer technical details of
our work.

\begin{figure}[t]
\begin{centering}
\includegraphics[width=1\columnwidth]{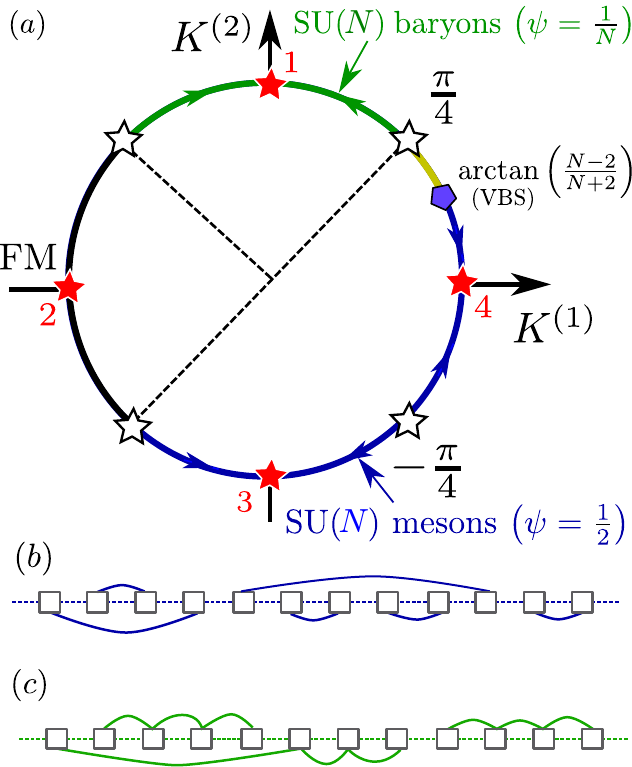}
\par\end{centering}
\caption{(a) Phase diagram of the strongly-disordered one-dimensional SO($N$)-symmetric
Hamiltonian of Eq.~\eqref{eq:Hamilt}. Points in the circle refer
to the angle $\tan\theta\equiv K_{i}^{\left(2\right)}/K_{i}^{\left(1\right)}$,
which is taken to be constant despite the randomness in $K_{i}^{\left(1,2\right)}$.
The blue and the green regions realize two distinct random-singlet
phases, \emph{both} with emergent SU($N$) symmetry. In the blue region,
SU($N$) singlets are built out of SO($N$) `spin' pairs {[}`mesons', shown
in panel (b){]}. The green region has SU($N$) singlets made of $kN$
`spins' (with $k=1,2,\ldots$) {[}`baryons', shown in panel (c){]}.
The arrows indicate the renormalization group flow. Red and white
stars represent stable and unstable fixed points, respectively. The
black (for any $N$) and the yellow (for even $N$) regions are not addressed
in this work. \label{fig:Full-phase}}
\end{figure}

\emph{Model and results.}\textendash{} The $N(N-1)/2$
SO($N$) generators {[}SO($N$) `spins'{]} will be denoted by $L^{ab}$,
with $a,b$ in the range $a=1,\ldots,N$ and $a<b$.~\footnote{We can adhere to this convention if we define $L^{ab}=-L^{ba}$ whenever
$a>b$.} We will take them in the defining representation, which is spanned
by a basis $\left|c\right\rangle $, $c=1,\ldots,N$. Each $L^{ab}$
generates rotations in the $ab$ plane. For $N=4$, for example, $L^{23}$
rotates a four-dimensional vector in the $\left(2,3\right)$ Cartesian
plane, while components 1 and 4 are kept fixed. In general,
\begin{equation}
iL^{ab}\left|c\right\rangle =\delta^{ac}\left|b\right\rangle -\delta^{bc}\left|a\right\rangle .\label{eq:generatoraction}
\end{equation}
The $L^{ab}$ operators obey the so($N$) Lie algebra 

\begin{equation}
\left[L^{ab},L^{cd}\right]=i\left(\delta^{bc}L^{ad}+\delta^{ad}L^{bc}-\delta^{ac}L^{bd}-\delta^{bd}L^{ac}\right),\label{eq:SON-algebra}
\end{equation}
with $\mbox{Tr}\left(L^{ab}L^{cd}\right)=2\delta^{ac}\delta^{bd}$. 

An SO($N$)-symmetric Hamiltonian can be built as a sum over pairs of SO($N$)
spins (although 3-site terms are possible, we do not consider them here). In the defining representation, the most general pair term
contains only bilinear and biquadratic terms~\cite{Tu_PhysRevB.78.094404,QuitoLopes2016PRB}.
In one dimension and considering only nearest-neighbor interactions
we have $H=\sum_{i}H_{i}$ where
\begin{equation}
H_{i}=J_{i}\mathbf{L}_{i}\cdot\mathbf{L}_{i+1}+D_{i}\left(\mathbf{L}_{i}\cdot\mathbf{L}_{i+1}\right)^{2},\label{eq:Hamilt0}
\end{equation}
where $\mathbf{L}_{i}\cdot\mathbf{L}_{i+1}=\sum_{a<b}L_{i}^{ab}L_{i+1}^{ab}$
and $J_{i},\,D_{i}$ are random couplings of $i$-th link. For later convenience, we will recast
$H$ in terms of the linear combinations $K_{i}^{\left(1\right)}=J_{i}-\frac{N-2}{2}D_{i}$
and $K_{i}^{\left(2\right)}=\frac{N-2}{2}D_{i}$, 

\begin{equation}
H_{i}=K_{i}^{\left(1\right)}\hat{O}_{i,i+1}^{\left(1\right)}+K_{i}^{\left(2\right)}\hat{O}_{i,i+1}^{\left(2\right)},\label{eq:Hamilt}
\end{equation}
where $\hat{O}_{i,i+1}^{\left(1\right)}=\mathbf{L}_{i}\cdot\mathbf{L}_{i+1}$
and $\hat{O}_{i,i+1}^{\left(2\right)}=\mathbf{L}_{i}\cdot\mathbf{L}_{i+1}+\frac{2}{N-2}\left(\mathbf{L}_{i}\cdot\mathbf{L}_{i+1}\right)^{2}$.

We choose a parametrization of Eq.~\eqref{eq:Hamilt}
in terms of the polar coordinates $\left(r_{i},\theta_{i}\right)$
in the $\left(K_{i}^{\left(1\right)},K_{i}^{\left(2\right)}\right)$
plane, so that $\tan\theta_{i}\equiv K_{i}^{\left(2\right)}/K_{i}^{\left(1\right)}$.
For simplicity, we  focus on random couplings $K_{i}^{\left(1\right)}$
and $K_{i}^{\left(2\right)}$ \emph{with a fixed ratio throughout the chain}, i.e., $\theta_{i}=\theta\ \forall i$ 
(the general case is discussed elsewhere~\cite{QuitoLopes2016PRB}).
In the regime of strong disorder, RSPs are found at low energies. The phase is determined by  $\theta$,
as displayed in a circle, see Fig.~\hyperref[fig:Full-phase]{\ref{fig:Full-phase}(a)}.
The basins of attraction, delineated by the colors and arrows,
are found via a strong-disorder renormalization group (SDRG) treatment~\cite{madasgupta,madasguptahu,fisher94-xxz,fishertransising2,igloi-review,*igloi-monthus-review2}.
The green and blue regions are both characterized by \emph{infinite
effective disorder} at long length scales~\cite{fisher94-xxz}.
More interestingly, both the blue and the green regions of Fig.~\hyperref[fig:Full-phase]{\ref{fig:Full-phase}(a)}
correspond to phases with emergent SU($N$) symmetry.

RSPs are characterized by a ground state formed by a collection of singlets. In the blue region, these random singlets
are formed by spin pairs {[}SO($N$) `mesons'{]}, as in the random Heisenberg
chain studied by Fisher~\cite{fisher94-xxz} {[}see Fig.~\hyperref[fig:Full-phase]{\ref{fig:Full-phase}(b)}{]}.
In such a phase, long bonds of length $L$ have strength of order
$\Omega\sim\exp\left(-L^{\psi_{M}}\right)$ with $\psi_{M}=1/2$.
Low-energy excitations correspond to breaking the longest bonds into
\emph{free} SO($N$) spins. At
temperature $T=\Omega$, bonds of length $L>L_{T}\sim\left|\ln T\right|^{1/\psi_{M}}$
are broken and the density of free spins is $n\left(T\right)\sim L_{T}^{-1}$.
Thermodynamic properties are then easily obtained: the spin linear
susceptibility follows from Curie's law $\left[\chi^{\left(1\right)}\right]^{-1}\sim T/n\left(T\right)\sim T\left|\ln T\right|^{1/\psi_{M}}$,
the entropy density is $s\left(T\right)\sim\left(\ln N\right)n\left(T\right)$
and the specific heat $c\left(T\right)=T\left(ds/dT\right)\sim\left|\ln T\right|^{-1-1/\psi_{M}}$.
A hallmark of the infinite effective disorder is the wide distribution
of correlation functions $C_{ij}=\left\langle \mathbf{L}_{i}\cdot\mathbf{L}_{j}\right\rangle $,
so that, at $T=0$, its average value is $C_{ij}^\text{av}\sim\left(-1\right)^{i-j}\left|i-j\right|^{-2}$
whereas the typical (i.e., most probable) one is $\left|C_{ij}^\text{typ}\right|\sim\exp\left(-\left(\left|i-j\right|/\xi\right)^{\psi_{M}}\right)$, 
where $\xi$ is a disorder-dependent length scale~\cite{getelina-hoyos-19}.

In the green region of Fig.~\hyperref[fig:Full-phase]{\ref{fig:Full-phase}(a)}, on the other hand,
the ground state consists of a collection of singlets formed out of
$kN$ ($k=1,2,\dots$) original SO($N$) spins {[}SO($N$) `baryons'{]} as depicted in Fig.~\hyperref[fig:Full-phase]{\ref{fig:Full-phase}(c)}.
The same relation between energy and length scales $\Omega\sim\exp\left(-L^{\psi_{B}}\right)$ holds, but now the exponent is $\psi_{B}=1/N$. 
Thermodynamic properties retain the same form described above but with $\psi_{M}\to\psi_{B}$.
Note that the structure of these RSPs is the same as the Damle-Huse
multicritical points~\cite{PhysRevB.66.104425,Damle2002}.

The emergent SU($N$) symmetry in each of these phases arises because,
as it turns out, the strongly entangled SO($N$) singlets, be they pairs
or $N$-tuples, are also SU($N$) singlets. Likewise, the original spins
into which these singlets are broken at energies above zero also transform
as SU($N$) spins. As these two types of objects ultimately determine
the low-energy properties, the latter will reflect this enhanced symmetry
group. For example, the susceptibilities of the SU($N$) operators (which
can be constructed from linear \emph{or bilinear} combinations of
the SO($N$) operators, as we will show) will also have the quoted behavior
with the same exponent in each phase. The same is true of the correlation
function distributions. These two types of phases and their properties
had been described before by two of us in disordered spin chains with
manifest SU($N$) symmetry~\cite{PhysRevB.70.180401}. Here, they are
realized asymptotically as emergent properties.

These are our main results. Their derivation relies on the application
of an elegant Lie algebra machinery to the SDRG. In what follows we
outline and motivate the results, relegating the full details to a
longer and more pedagogic exposition~\cite{QuitoLopes2016PRB}.

\emph{SDRG details.}\textendash{} The SDRG method is based on an iterative
removal of degrees of freedom in real space following an energy hierarchy
dictated by the largest local 2-site gap. Each iteration step consists
of (i) the decimation of the pair with largest gap $\Omega$ by a
projection of its Hilbert space onto its ground multiplet and (ii)
the renormalization of the remaining couplings between this sub-space
to the adjacent spins using perturbation theory. When applied sequentially,
this process translates into a flow of the distribution of coupling
constants. While the form of the Hamiltonian and the connectivity
of the chain is preserved, new multiplets belonging
to any one of the anti-symmetric SO($N$) representations appear throughout
the flow. As a consequence, the full characterization of the phases involves a flow of representation distributions.

Using Eq.~\eqref{eq:Hamilt}, the decimation rules can be written
in closed form~\cite{QuitoLopes2016PRB}. Crucially, the decimations of the angles $\theta_{i}$ do not
involve the radial variables $r_{i}$. Suppose the largest gap occurs
between spins 2 and 3. If the ground multiplet of $H_{2,3}$ is not
a singlet, it belongs to one of the $\mathrm{int}\left(N/2\right)$
anti-symmetric representation of SO($N$), and spins $2$ and $3$ are
replaced by a new spin in that representation. The couplings in links
1 and 3 are renormalized according to

\begin{equation}
\tan\tilde{\theta}_{1,3}=\pm\tan\theta_{1,3}.\label{eq:deci1ord}
\end{equation}
The choice of sign is determined by the representations being decimated
as well as their ground state multiplet~\cite{QuitoLopes2016PRB}.
If the ground multiplet of $H_{2,3}$ is a singlet, spins $2$ and
$3$ are effectively removed. In this case, a new coupling between
spins 1 and 4 is created with~\cite{QuitoLopes2016PRB}
\begin{equation}
\tan\tilde{\theta}=-\left(\frac{N+2}{N}\right)\frac{\frac{N-2}{N+2}-\tan\theta_{2}}{1-\tan\theta_{2}}\tan\theta_{1}\tan\theta_{3}.\label{eq:deci2ord}
\end{equation}

In the blue mesonic region of Fig.~\hyperref[fig:Full-phase]{\ref{fig:Full-phase}(a)}, the
ground multiplets are always singlets and it follows trivially from
Eq.~\eqref{eq:deci2ord} that $\theta=0$ and $-\pi/2$ (points 3
and 4 of the Figure) and $\theta=-\pi/4$ are fixed points of the
flow. The same equation can be used to show that points 3 and 4 are
stable whereas $\theta=-\pi/4$ is unstable.

\begin{figure}[t]
\begin{centering}
\includegraphics[width=1\columnwidth]{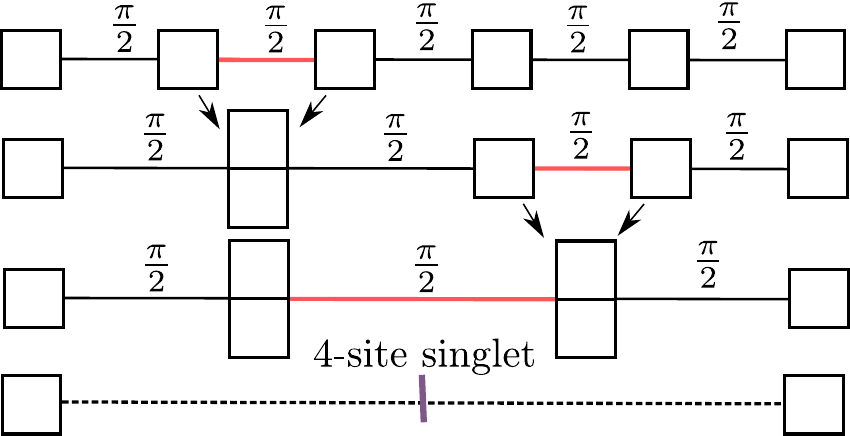}
\par\end{centering}
\caption{\label{fig:baryonic}An example of singlet formation for SO(4) at
the fixed point 1 of Fig.~\hyperref[fig:Full-phase]{\ref{fig:Full-phase}(a)},
with $\theta_{i}$ indicated on each bond.}
\end{figure}

In the green baryonic region of Fig.~\hyperref[fig:Full-phase]{\ref{fig:Full-phase}(a)} both
types of decimations occur and the analysis is more involved. The
pair of angles $\theta=\pm\pi/2$ taken together are fixed points
and singlets are formed out of $kN\text{ }\left(k=1,2,\ldots\right)$
of SO($N$) spins. There are several paths by which this can happen and
an illustrative example is shown in Fig.~\ref{fig:baryonic} for
SO(4). In this case, the RG flow involves two anti-symmetric representations
depicted by Young tableaux with 1 or 2 stacked boxes. Note how the
angle can switch back and forth from $\pi/2$ to $-\pi/2$ depending
on the representations involved. This is the fixed point 1 in Fig.~\hyperref[fig:Full-phase]{\ref{fig:Full-phase}(a)}.
A stability analysis shows that point 1 is a stable fixed point. Similarly,
the extremities of the green region $\theta=\pi/4$ and $\theta=3\pi/4$
are unstable fixed points since, crucially, they lead to Hamiltonians
with exact SU($N$) symmetry and this symmetry is preserved by the SDRG
flow.

We now show that the SO($N$)-symmetric Hamiltonian of Eq.~\eqref{eq:Hamilt}
can be viewed as an SU($N$)-anisotropic problem. The $N^{2}-1$
generators $\left\{ \Lambda_{i}\right\} $ of the fundamental representation
of the SU($N$) group are traceless Hermitian matrices, normalized as
$\mbox{Tr}\left[\Lambda_{i}^{\left(a\right)}\Lambda_{j}^{\left(b\right)}\right]=2\delta^{ab}\delta_{ij}$.
We can break this set in a subset of $N\left(N-1\right)/2$
\textit{\emph{purely imaginary}}\emph{ }\textit{\emph{anti-symmetric}}
matrices, the generators of SO($N$), and another subset of $N\left(N+1\right)/2-1$
\textit{\emph{real}}\emph{ }\textit{\emph{traceless symmetric}} ones,
which are SO($N$) second-rank tensors {[}see the form of
$\hat{O}_{i,i+1}^{\left(2\right)}$ after Eq.~\eqref{eq:Hamilt}{]}.
The Hamiltonian \eqref{eq:Hamilt} is then equivalent to an SU($N$)-anisotropic
Hamiltonian,

\begin{eqnarray}
H_{i} & = & K_{i}^{\left(1\right)}\sum_{a=1}^{d_{\text{SO}(N)}}\Lambda_{i}^{\left(a\right)}\Lambda_{i+1}^{\left(a\right)}+K_{i}^{\left(2\right)}\!\!\!\!\sum_{a=d_{\text{SO}(N)}+1}^{N^{2}-1}\!\!\!\!\Lambda_{i}^{\left(a\right)}\Lambda_{i+1}^{\left(a\right)},\,\,\label{eq:SUN_aniso}
\end{eqnarray}
with $d_{\text{SO}(N)}=\frac{N\left(N-1\right)}{2}.$ We can immediately
find the expected SU($N$)-symmetric points: $K_{i}^{\left(1\right)}=\pm K_{i}^{\left(2\right)}$.
That the choice with a minus sign is also SU($N$)-symmetric can be seen
from the transformation $\Lambda_{i}^{\left(a\right)}\to-\Lambda_{i}^{\left(a\right)*}\equiv\tilde{\Lambda}_{i}^{\left(a\right)}$
on every other chain site, which changes an SU($N$) representation into
its conjugate and absorbs the minus sign. This case corresponds to
having SU($N$) (anti-) fundamental representations on odd (even) sites.

The location of these angular fixed points sets the topology of the
flow, as shown by the arrows in Fig.~\hyperref[fig:Full-phase]{\ref{fig:Full-phase}(a)}. Although
the $\theta$-distribution starts as a delta function, it broadens
under the SDRG flow. The existence of the stable fixed points, however,
forces the distribution to narrow back down to a delta function at
one of the points 1, 3 or 4. Point 2 and its associated black region
are outside the scope of this paper as symmetric representations of
SO($N$) are generated by the flow. The yellow region between the generalized
AKLT point $\theta_\text{VBS}=\arctan\left[\left(N-2\right)/\left(N+2\right)\right]$
(blue pentagon)~\cite{PhysRevLett.59.799} and $\pi/4$ flows to
the fixed point 4 for odd $N$. For even $N$, the procedure becomes ill-defined in this region, and our method cannot be applied~\cite{QuitoLopes2016PRB}.

The renormalization of radial variables depends explicitly on the
representations being decimated as well as the effective ones being
introduced. A systematic derivation of such rules will be given elsewhere~\cite{QuitoLopes2016PRB},
but up to pre-factors, the rules are the ones derived in Ref.~\onlinecite{Damle2002}.
The distribution of $r_{i}$ broadens without limit and flows to an
infinite disorder form given by $P\left(r\right)\sim r^{\alpha_{i}\left(\Omega\right)-1}$.
Here $\alpha_{i}\left(\Omega\right)=\left(\psi_{i}^{-1}-1\right)/\left|\ln\Omega\right|$,
$i=B$ or $M$ in the green or blue region, respectively, and $\Omega$
is the decreasing cutoff of the distribution~\cite{fisher94-xxz,fishertransising2,Quito_PhysRevLett.115.167201,QuitoLopes2016PRB}.

In the blue region, adjacent spins always form a singlet and no other
representation appears in the flow. The ground state structure is
shown in Fig.~\hyperref[fig:Full-phase]{\ref{fig:Full-phase}(b)}. In contrast, in the green
region decimations with ground multiplets belonging to any one of
the $\mathrm{int}\left(N/2\right)$ antisymmetric representations
of SO($N$) are generated. After an initial transient,
each one of them is equally populated in the renormalized system.~\footnote{With one exception: for even N, the self-conjugate representation is half as likely as any of the others.} A singlet only forms out of $kN\text{ }\left(k=1,2,\ldots\right)$
SO($N$) spins, leading to the ground state structure in Fig.~\hyperref[fig:Full-phase]{\ref{fig:Full-phase}(c)}.
The different singlet structures lead to different physical properties
at finite energies, as discussed above. The apparently intricate combinations
leading to singlet formation out of $kN$ SO($N$) spins can
be easily understood at the exact SU($N$) point $\theta=\pm\pi/4$:
only with $kN$ SU($N$) fundamentals can one form an SU($N$)
singlet~\cite{PhysRevB.70.180401}. The stable fixed points that
attract the flow are adiabatically connected to these SU($N$) points
and have the same ground state structure.

The emergent SU($N$) symmetry, as mentioned, relies on the fact that
free spins and frozen singlets, the building blocks of the renormalized
system, transform as SU($N$) fundamentals and singlets, respectively.
If we now recall that some of the SU($N$) generators $\Lambda_{i}^{\left(a\right)}$
with $a\in\left[N\left(N-1\right)/2+1,N^{2}-1\right]$
are actually 2nd-rank SO($N$) tensors (see Eqs.~\eqref{eq:Hamilt}
and \eqref{eq:SUN_aniso}), it follows that susceptibilities and correlation
functions built with these quadratic SO($N$) operators are governed
by the same power laws as those of the SO($N$) generators. Measuring
SO($N$) susceptibilities may sound as a challenging task. Yet, we point
that this can be envisaged at least for the case of $N=3$. In this
case these susceptibilities are just regular magnetic susceptibilities
for spin-1 operators~\cite{Quito_PhysRevLett.115.167201}. The susceptibilities
for 2nd rank operators in this case are nothing but quadrupolar susceptibilities;
protocols for their measurements have recently been proposed at least
in two dimensions by considering cross responses between magnetic
probes and strain~\cite{Patri_Multipolar}.

\emph{Conclusions.}\textemdash{} Our study of random SO($N$)-symmetric
chains unveils a unified mechanism of symmetry emergence in a large
and diverse set of realizable physical situations. Some possible realizations
had been previously studied ($N=2,3$) but new ones ($N=4,6$)
are here introduced. Crucial to the mechanism is the existence of
explicit SU($N$)-symmetric points in the parameter space whose ground
states are adiabatically connected (no local-gap closing) to those
of a finite region: symmetry emergence requires no fine tuning. Disorder
is the ingredient responsible for filtering, from the set of SO($N$)
representations, those which find correspondence in the SU($N$) group.

\emph{Acknowledgment - }We thank Gabe Aeppli for discussions. VLQ
and PLSL contributed equally to this work. VLQ acknowledges financial
support from the National High Magnetic
Field Laboratory through grant DMR-1157490, the State of Florida and the
Aspen Center for Physics, supported by NSF grant PHY-1607611, for
hospitality. PLSL is supported by the Canada First Research Excellence
Fund. JAH and EM acknowledge financial support from FAPESP, CNPq and
Capes.

\emph{Author contributions - }V.L.Q. and P.L.S.L. performed the necessary calculations, with inputs of E.M. and J.A.H. All the authors participated in the discussions of the results. E.M. and J.A.H. suggested the connections with higher-order susceptibilities. The manuscript was written with the equal collaboration of all authors.

\bibliographystyle{apsrev4-1}
\bibliography{bibliog_son}

\end{document}